\documentclass[intlimits,twoside,a4paper]{article}
\usepackage{amsmath,amssymb}
\usepackage{graphicx}
\usepackage[T2A]{fontenc}
\usepackage[cp1251]{inputenc}

\usepackage[eqsecnum]{cmpj}

\issue{2011}{14}{3}{33602}

\doinumber{10.5488/CMP.14.33602}

\title[Primitive models of molecular ionic liquids]{Phase behaviour and
dynamics in primitive models of molecular ionic liquids}

\author[G.C. Ganzenm{\"u}ller, P.J. Camp]{G.C.
Ganzenm{\"u}ller\refaddr{label1}\thanks{E-mail: georg.ganzenmueller@emi.fraunhofer.de}\,, P.J. Camp\refaddr{label2}\thanks{E-mail: philip.camp@ed.ac.uk}}

\addresses{\addr{label1} Fraunhofer Ernst Mach Institute for High-Speed
Dynamics, 4~Eckerstrasse, 79104~Freiburg, Germany
\addr{label2} School of Chemistry,
The University of Edinburgh, West Mains Road, Edinburgh~EH9~3JJ, United
Kingdom}

\authorcopyright{G.C. Ganzenm{\"u}ller, P.J. Camp, 2011}
\date{Received March 4, 2011}

\begin{document}

\maketitle

\begin{abstract}

The phase behaviour and dynamics of molecular ionic liquids are studied
using primitive models and extensive computer simulations. The models
account for size disparity between cation and anion, charge location on
the cation, and cation-shape anisotropy, which are all prominent
features of important materials such as room-temperature ionic liquids.
The vapour-liquid phase diagrams are determined using high-precision
Monte Carlo simulations, setting the scene for in-depth studies of ion
dynamics in the liquid state. Molecular dynamics simulations are used to
explore the structure, single-particle translational and rotational
autocorrelation functions, cation orientational autocorrelations, self
diffusion, viscosity, and frequency-dependent conductivity. The results
reveal some of the molecular-scale mechanisms for charge transport,
involving molecular translation, rotation, and association.
\pacs 61.20.Ja, 64.70.F-, 66.10.C-, 66.10.Ed, 66.20.-d

\end{abstract}

\section{Introduction}

Molecular ionic liquids are low-melting point compounds made up of
molecular cations and anions. Currently, the most important examples are
room-temperature ionic liquids (RTILs), which have very low vapour
pressures, solvate a wide range of polar and non-polar solutes, and have
been shown to catalyse chemical transformations when used as reaction
media~\cite{Welton:1999/a}. Typical cations (such as dialkylimidazolium)
may exhibit both ionic and non-ionic characteristics. For instance,
solutions of RTILs and water lead to the formation of micellar phases~\cite{Wang:2005/a,CanongiaLopes:2006/a,CanongiaLopes:2006/b,Triolo:2007/a,%
Triolo:2009/a}. In terms of transport properties, RTILs are generally
considered quite viscous and the diffusion constants are correspondingly
low as compared to those in non-ionic molecular liquids. Experimentally,
measurements of the frequency-dependent dielectric response yield
insights on microscopic motions that lead to a change of polarization
and, through the fundamental link between dielectric response and
conductivity~\cite{Schroder:2008/a,Schroder:2009/a}, to the transport of
charge. The dielectric spectra of many common RTILs have been measured~\cite{Schrodle:2006/a,Tokuda:2004/a,Tokuda:2005/a,Tokuda:2006/a,%
Hayamizu:2010/a,Mizoshiri:2010/a}. Of particular note, Weing{\"a}rtner
and colleagues found evidence for a contribution to the dielectric
spectra of 1-alkyl-3-methylimidazolium salts likely arising from cation
rotations~\cite{Schroder:2007/a,Wulf:2007/a,Huang:2010/a}. An analysis
of dielectric spectra and nuclear-magnetic relaxation measurements
highlighted motions on the order of tens of picoseconds (corresponding
to real frequencies $\nu \sim 10^{11}$~{Hz}). Computer-simulation
results have been used to resolve the different contributions to the
dielectric spectrum, with the conclusion that the dielectric relaxation
arising from molecular translations is faster than that from molecular
rotations~\cite{Schroder:2007/a}.

Atomistic simulations have yielded invaluable insights on the important
molecular interactions and correlations in specific systems~\cite{DelPopolo:2004/a,CanongiaLopes:2004/a}; there are far too many
simulation studies to mention here, but a review by Maginn highlights
the main achievements and outstanding challenges~\cite{Maginn:2009/a}.
Low-frequency dynamics can be studied in atomistic simulations~\cite{Shim:2008/a} but with some difficulty. The primary problem is that
to calculate transport coefficients or frequency-dependent response
functions requires accurate calculations of appropriate time-correlation
functions, but the rate of decay of these functions is
characteristically long in RTILs and it is a challenge to calculate the
long-time tails with acceptable signal-to-noise ratios~\cite{Maginn:2009/a}. Free-energy calculations and the construction of
phase diagrams are other areas where the complexity of atomistic models
may push the calculations out of reach. Therefore, to achieve a more
comprehensive survey of dynamics over broad timescales~-- as well as the
structure and phase behaviour~-- one might consider simplified models in
which certain molecular characteristics can be tuned at will.

Clearly, the main molecular characteristics of RTILs are that the ions
are of different sizes, the charges are distributed unevenly over the
ions (especially in typical cations), and the cations are normally far
from spherical. Typical anions such as $\mbox{Cl}^{-}$,
$\mbox{BF}_{4}^{-}$, or $\mbox{PF}_{6}^{-}$, may be considered spherical
in a first approximation, but cations such as dialkylimidazolium or
dialkylpiperidinium are heterocyclic species. In the same way that
a fluid of charged hard spheres, the restricted primitive model (RPM), serves
as a basic model of simple molten salts, certain extensions to allow for
the aforementioned molecular characteristics can be studied to gain
insight on molecular ionic liquids. Malvaldi and Chiappe studied the
dynamics in primitive models of RTILs consisting of dumbbell cations
(each made up of two soft spheres, with one or both carrying a central
charge) and simple soft-sphere anions~\cite{Malvaldi:2008/a}. It was
shown that, even with such a simple model, certain experimental
observations could be reproduced and rationalised in microscopic terms.
For instance, in some imidazolium RTILs, the diffusion constant of the
cation is greater than that of the anion~\cite{Umecky:2005/a,Kanakubo:2007/a}; atomistic simulations show that
the cation diffuses preferentially by moving in the direction of the
carbon in the 2 position on the five-membered ring (in between the
nitrogens at the 1 and 3 positions)~\cite{Urahata:2005/a}. This
translational anisotropy is captured to some degree by Malvaldi and
Chiappe's dumbbell model where the charge is distributed equally between
the two halves; at low temperature and high density, the cation diffuses
faster than the anion. Spohr and Patey have performed a highly
systematic and comprehensive survey of the microscopic structure and
dynamics in models consisting of a large spherical cation carrying a
point charge displaced from the centre, and a simple soft-sphere anion.
By controlling the location of the cation charge and cation-anion size
disparity, a number of experimental trends could be rationalised on the
basis of ion-ion correlations~\cite{Spohr:2008/a,Spohr:2009/a,Spohr:2010/a}. Specifically, many of the
experimentally observed dependences of shear viscosity, conductivity,
and diffusion constants on relative ion size, molecular charge
distribution, and temperature could be mimicked with the simple models.
Recent work on such models has even captured the effects of a polar
solvent/impurity, such as water~\cite{Spohr:2010/b}.

Regarding phase behaviour, the available thermodynamic data on RTILs
show some interesting trends. For instance, the vapour-liquid critical
temperature ($T_{\rm c}$)~-- although difficult to measure directly in
experiments because of molecular decomposition at high temperatures~--
appears to decrease with increasing molecular weight of the alkyl chains
on the imidazolium cations, despite the growing van der Waals
interactions favouring an increase in $T_{\rm c}$~\cite{Rebelo:2005/a}.
Mart{\'i}n-Betancourt  et al. constructed coarse-grained models in
which the cations are represented by charged hard spherocylinders, and
performed high-precision Monte Carlo (MC) simulations to determine the
coexistence envelopes and critical points~\cite{Martin-Betancourt:2009/a}. The critical temperature and density
both decrease with increasing cation elongation, in qualitative
correspondence with experimental data, due to the growing entropic role
played by the steric bulk of the cation. Schr{\"o}er and Vale surveyed
fluid-fluid phase separations in solutions of imidazolium salts in a
variety of polar and non-polar solvents, and showed that the phase
diagrams could be analysed in terms of those for charged hard spheres
with size asymmetry~\cite{Schroer:2009/a}. The main conclusion here was
that the phase separation is mainly driven by Coulombic interactions,
and that steric and dispersion interactions (which can be controlled
systematically with substituents on the cations) modify the critical
parameters. Models such as size-asymmetric charged hard spheres can be
studied to yield systematic variations in critical parameters with
cation size; integral-equation studies point to some non-trivial effects~\cite{Kalyuzhnyi:2004/a,Kalyuzhnyi:2005/a}.

In this work, primitive models of molecular ionic liquids are
constructed and their phase behaviour and dynamical properties are
examined. The models are chosen to reflect some of the key molecular
characteristics, such as a size disparity between cation and anion, the
location of the positive charge on the molecular cation, and the shape
anisotropy of the cation. By constructing simple models possessing each
of these characteristics, the essential relationships between molecular
properties and dynamical behaviour can be identified. First, the
vapour-liquid phase diagrams are computed using high-precision MC
simulations. Then, molecular dynamics (MD) simulations of the dense
liquid are performed in order to evaluate the microscopic structure and
the transport coefficients~-- diffusion constants, viscosity, and
conductivity. The dependences of these properties on molecular
architecture and temperature are briefly discussed. Finally, the
frequency-dependent conductivity is considered and the contributions
from cation translations and rotations are resolved by comparing the
conductivity spectrum with the spectra of the translational and
rotational-velocity autocorrelation functions; the orientational
correlations of the cations are also considered. The possible relevance
of these results to experimental measurements of the dielectric response
is discussed.

This paper is organised as follows. In section~\ref{sec:models}, the
molecular models and parameters are defined. Monte Carlo simulations of
phase behaviour are presented in section~\ref{sec:phase}, focusing on
the vapour-liquid coexistence envelope and the associated critical
point. Molecular dynamics simulations of microscopic structure and
(frequency-dependent) transport properties are discussed in section~\ref{sec:strdyn}. Section~\ref{sec:conclusions} concludes the paper.

\section{Models} \label{sec:models}


\begin{figure}[!b]
\centering
\includegraphics[width=11.0cm]{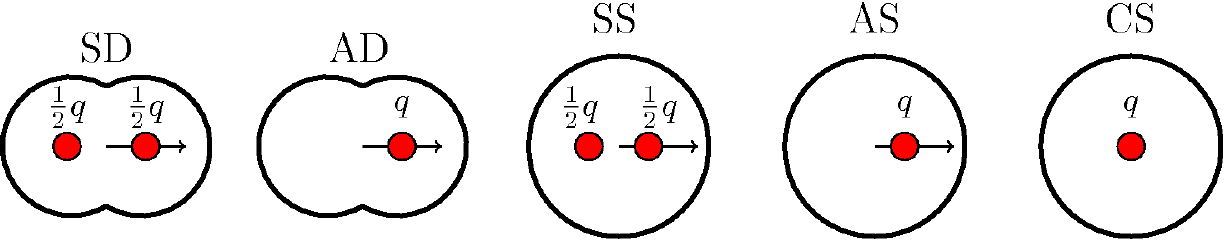}
\caption{\label{fig:models} Models of primitive liquids distinguished by
the model for the cation: symmetric dumbbell (SD); asymmetric dumbbell
(AD); symmetric sphere (SS); asymmetric sphere (AS); centred sphere
(CS). In each case, the anion is modelled as a single charged LJ sphere
with diameter $\sigma$, charge $-q$, and mass $2m$. The SD and AD
cations are each modelled with two LJ spheres of diameter $\sigma$ and
mass $m$, fused together at a distance $d=\sigma/\sqrt{2}$: in the SD
model, each sphere carries a charge $+q/2$; and in the AD model, only
one sphere carries the charge $+q$. The SS, AS, and CS cations are each
modelled by a charged LJ sphere with diameter $\sigma_{+}=1.3014\sigma$
and mass $2m$: the SS cation carries two charges $+q/2$ displaced
symmetrically from the centre by $0.1507\sigma$; the AS cation has a
single charge $+q$ located $0.1507\sigma$ from the centre; the CS cation
has its charge $+q$ at the centre. With these choices, the mass of each
cation is $2m$, the shortest distance from the positive charges(s) to
the edge is $\sigma/2$ (except for the CS model), and the excluded
volume of each cation is the same. For the SD, AD, SS, and AS models,
arrows indicate the unit vectors used to define the orientational
autocorrelation function in section~\ref{sec:dynamics}.}
\end{figure}

The primitive models are presented in figure~\ref{fig:models}. In each
case the anion is modelled as a single charged Lennard-Jones (LJ)
sphere with energy parameter $\epsilon$, diameter $\sigma$, charge $-q$,
and mass $2m$. In the symmetric dumbbell (SD) and asymmetric dumbbell
(AD) models, each cation is modelled with two LJ spheres (with LJ
parameters $\epsilon$ and $\sigma$, and mass $m$) fused together at a
distance $d=\sigma/\sqrt{2}$: in the SD model, each sphere carries a
central charge of $+q/2$; in the AD model, only one sphere carries a
central charge $+q$. In the corresponding symmetric-sphere (SS) and
asymmetric-sphere (AS) models, each cation is formed from a single large
LJ sphere with diameter $\sigma_{+}$\,, energy parameter $\epsilon$, and
mass $2m$. The diameter $\sigma_{+}$ is chosen so that the excluded
volume $4\pi\sigma_{+}^{3}/3$ is equal to that of the dumbbells defined
above. The excluded volume of two dumbbells with sphere diameter
$\sigma$ and separation $d$ is given approximately by~\cite{Boublik:1977/a}
\begin{equation}
\frac{\pi\sigma^{3}}{6}
\left[ 8
      +12\left(\frac{d}{\sigma}\right)
      +3\left(\frac{d}{\sigma}\right)^{2}
      -\left(\frac{d}{\sigma}\right)^{3}
\right].
\end{equation}
With $d=\sigma/\sqrt{2}$\,, the effective cation diameter is $\sigma_{+}
\simeq 1.3014\sigma$. The charges on the cations are placed so that the
shortest distance to the edge is always $\sigma/2$: in the SS model, two
charges $+q/2$ are displaced symmetrically from the centre by a distance
$(\sigma_{+}-\sigma)/2=0.1507\sigma$; in the AS model, one charge $+q$
is placed $0.1507\sigma$ from the centre. Finally, in the centred-sphere
(CS) model, the cation consists of a single charged LJ sphere with
diameter $\sigma_{+}=1.3014\sigma$, energy parameter $\epsilon$, mass
$2m$, and central charge $+q$. With this set of parameters, the
excluded volumes for all cations are equal, the shortest distances from
the charges to the edges are all equal to $\sigma/2$ (except for the CS
model), the charges on the cations and anions are $\pm q$, and the
cations and anions have equal masses of $2m$. The interactions between
ions are given by a sum of LJ and Coulombic (C) site-site interactions.
The LJ interaction between two spheres $i$ and $j$ is given by
\begin{equation}
u_{ij}^{\rm LJ}
= 4\epsilon\left[ \left(\frac{\sigma_{ij}}{r_{ij}}\right)^{12}
                 -\left(\frac{\sigma_{ij}}{r_{ij}}\right)^{6}
            \right],
\label{eqn:lj}
\end{equation}
where $\sigma_{ij}=(\sigma_{i}+\sigma_{j})/2$, $r_{ij}$ is the
centre-centre distance between spheres, and for simplicity the energy
parameter $\epsilon$ is the same for all interactions. The charge-charge
interaction is
\begin{equation}
u_{ij}^{\rm C} = \frac{q_{i}q_{j}}{{\cal E}s_{ij}}\,,
\label{eqn:uc}
\end{equation}
where $q_{i}$ is the charge on particle $i$, $s_{ij}$ is the distance
between charges $i$ and $j$, and ${\cal E}=4\pi\varepsilon_{0}$ in which
$\varepsilon_{0}$ is the dielectric permittivity of the vacuum. The
reduced charge is defined as $q^{*}=\sqrt{q^{2}/{\cal
E}\sigma\epsilon}$; in this work, there is given a fixed value of
$q^{*}=7$. This is appropriate for monovalent ions ($q=e$) and realistic
choices for the LJ energy parameter $\epsilon/k_{\rm B} \simeq
600~\mbox{K}$ and the LJ diameter $\sigma \simeq 5~\mbox{\AA}$. For each
of the dumbbell models, the reduced moment of inertia of the cation is
$I^{*}=I/m\sigma^{2}=0.25$; the same value is used for each of the
spherical-cation models. Other reduced units are defined using
$\epsilon$, $\sigma$, and $m$: the reduced temperature is $T^{*}=k_{\rm
B}T/\epsilon$, where $k_{\rm B}$ is Boltzmann's constant; the reduced
ion-pair concentration $\rho^{*}=N\sigma^{3}/V$, where $N$ is the number
of ion pairs and $V$ is the volume; the reduced pressure
$P^{*}=P\sigma^{3}/\epsilon$; and the reduced time $t^{*}=t/\tau$ where
$\tau=\sqrt{m\sigma^{2}/\epsilon}$\,. With the typical values of
$\epsilon$ and $\sigma$ given above, along with a characteristic salt
molecular weight of $4m \simeq 300~\mbox{g}~\mbox{mol}^{-1}$, the basic
unit of time is $\tau \simeq 1.94~\mbox{ps}$.

The phase diagrams of the model systems were determined using MC
simulations; the simulation method and results are presented in section~\ref{sec:phase}. Subsequently, the dynamical properties were determined
using MD simulations; the computational details and results are
presented in section~\ref{sec:strdyn}. In all cases, the systems were
simulated in cubic boxes of side $L$ and with periodic boundary
conditions applied. The LJ interactions were truncated and shifted at
$L/2$, and the Coulombic interactions were handled using the Ewald
summation with conducting periodic boundary conditions~\cite{Allen:1987/a}.

\section{Phase behaviour} \label{sec:phase}

\subsection{Monte Carlo simulations} \label{sec:mc}

To identify the liquid region of the phase diagram and to focus the
subsequent dynamical studies, vapour-liquid coexistence envelopes were
determined using a MC technique. Wang-Landau simulations were performed
in the grand-canonical ensemble according to the GCMCWL scheme described
in reference~\cite{PJC:2007/e}. Essentially, this is a flat-histogram
sampling method that enables an iterative determination of the canonical
partition function $Q(N,V,T)$ as a function of $N$ at fixed $V$ and $T$.
The conditions for phase coexistence are then easily determined using an
equal-area criterion applied to the bimodal particle-number distribution
$p(N) \propto z^{N}Q(N,V,T)$, where $z$ is the activity. Full details
are given in reference~\cite{PJC:2007/e}. The current simulations were
carried out in cubic boxes with sides $L=11.75\sigma$ and $L=13\sigma$
for dumbbell and sphere models, respectively. The GCMCWL technique has
been successfully applied  to a number of `tough' systems where a large
degree of clustering is anticipated near the coexistence region,
including charged soft spheres~\cite{PJC:2007/e} and dipolar spheres~\cite{PJC:2008/c}.

\subsection{Vapour-liquid coexistence envelopes} \label{sec:vl}

\begin{figure}[h!]
\centering
\includegraphics[width=8cm]{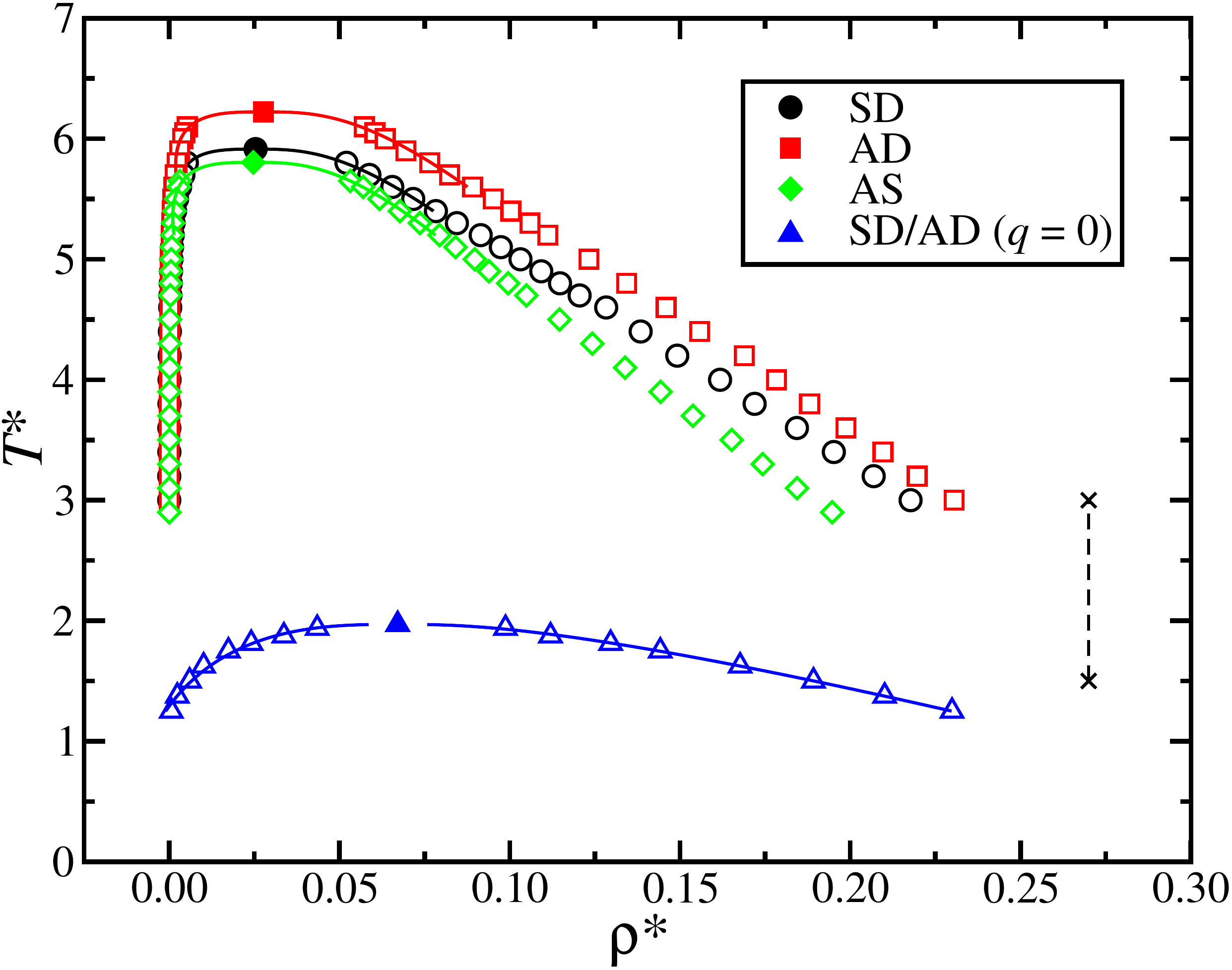}
\caption{\label{fig:vl} (Color on-line) Phase diagrams of selected primitive models of
ionic fluids: (circles) SD model; (squares) AD model; (diamonds) AS
model; (triangles) SD/AD model with $q=0$. Open symbols indicate
sub-critical coexistence points, filled symbols indicate the critical
points, and the lines show the fits from the scaling law in equation~(\ref{eqn:scaling}). The dashed line indicates the temperature range
along the $\rho^{*}=0.27$ isochore studied using molecular dynamics
simulations.}
\end{figure}

Vapour-liquid coexistence envelopes for the SD, AD, and AS systems are
shown in figure~\ref{fig:vl}. In order to assess the effects of the
Coulombic interactions, results are also shown for an equimolar mixture
of {\em uncharged} dumbbells and LJ spheres (the SD/AD model with
$q=0$). The critical parameters were estimated by fitting the simple
scaling law
\begin{equation}
\rho_{\pm} = \rho_{\rm c} + At \pm Bt^\beta
\label{eqn:scaling}
\end{equation}
to near-critical coexistence densities in the vapour ($\rho_{-}$) and
liquid ($\rho_{+}$) phases, where $\rho_{\rm c}$ is the critical
density, $t=1-T/T_{\rm c}$\,, and $\beta=0.3265$ is the 3D Ising
order-parameter exponent. Ionic criticality is known to be Ising-like~\cite{Luijten:2002/b}. The fitted scaling laws are shown in figure~\ref{fig:vl}.


The critical parameters are given in table~\ref{tab:critical}. By
comparing the charged and uncharged cases, it is clear that the
Coulombic interactions result in a large increase in the critical
temperature and a substantial decrease in the critical density. A low
critical density is one characteristic of an ionic critical point~\cite{Fisher:1996/a}, and with the current parameters, the Coulombic
interactions are very strong. To compare the current results with those
for the simple case of charged hard spheres~-- the restricted primitive
model (RPM)~-- table~\ref{tab:critical} also shows the packing fraction
$\eta$ at the critical point. For the systems studied here, the volumes
of the ions are estimated by approximating them as hard particles. For
the SD and AD models, the volume of a dumbbell is
\begin{equation}
v(d) = \frac{\pi\sigma^{3}}{6}
       \left[   1
              + \frac{3}{2}\left(\frac{d}{\sigma}\right)
              - \frac{1}{2}\left(\frac{d}{\sigma}\right)^{3}
    \right].
\label{eqn:vdb}
\end{equation}
The key point is that $\eta_{\rm c} \simeq 0.04$ for all charged models,
including the RPM. The comparison of critical temperatures is not that
straightforward. The RPM critical temperature is $k_{\rm B}T_{\rm
c}=0.05069(2)q^{2}/{\cal E}\sigma$~\cite{Luijten:2002/b}, where
$-q^{2}/{\cal E}\sigma$ is the Coulombic interaction between a cation
and an anion at contact. For the primitive models studied here, the
minimum in the interaction potential between two oppositely charged LJ
spheres is $r_{\rm min} \simeq 0.9556\sigma$ and $u_{\rm min} =
-49.63\epsilon$. The reduced critical temperatures are all in the region
of $T_{\rm c}^{*} = 6$ and so the corresponding `ionic' temperature is
$k_{\rm B}T_{\rm c}/|u_{\rm min}| \simeq 0.12$. This is considerably
higher than the RPM value, reflecting the soft cores of the particles
and the additional LJ attractions. Model AD has the highest critical
temperature, reflecting the fact that there are a large number of
cation-anion arrangements where the charge separation is about $1\sigma$
and the Coulombic interaction is strong. The number of such arrangements
is reduced when the cation charge is split between two centres, as in
the SD model, because now the anion has to be in the right place to
satisfy two site-site interactions. For model AS, only one arrangement
of the anion and cation gives a charge-charge separation of about
$1\sigma$, and so the critical temperature is relatively low.


\begin{table}[t]
\caption{\label{tab:critical} Vapour-liquid critical parameters for
selected primitive models of ionic fluids (including the RPM~\cite{Luijten:2002/b}) and for an uncharged system of dumbbells and LJ
spheres (SD/AD model with $q=0$). $T_{\rm c}^{*}=k_{\rm B}T_{\rm
c}/\epsilon$ is the reduced critical temperature, $\rho_{\rm
c}^{*}=\rho_{\rm c}\sigma^{3}$ is the reduced critical ion-pair density,
and $\eta_{\rm c}$ is the critical packing fraction calculated from
$\rho_{\rm c}^{*}$ assuming hard particles of dimension $\sigma$. The
figures in brackets are the estimated uncertainties in the final
digits.}
\vspace{2ex}
\centering
\begin{tabular}{|l|l|l|l|}
\hline
Model & $T_{\rm c}^{*}$ & $\rho_{\rm c}^{*}$ & $\eta_{\rm c}$ \\ \hline\hline
SD                        & $5.91(2)$  & $0.0255(1)$ & $0.0385(3)$ \\
AD                        & $6.22(2)$  & $0.0281(2)$ & $0.0424(3)$ \\
AS                        & $5.80(2)$  & $0.0247(1)$ & $0.0414(2)$ \\
SD/AD ($q=0$)             & $1.970(8)$ & $0.0671(6)$ & $0.1013(9)$ \\
RPM~\cite{Luijten:2002/b} & $0.05069(2){q^{*}}^{2}$  & $0.0395(1)$ & $0.0414(1)$ \\
\hline
\end{tabular}
\end{table}

Knowledge of the vapour-liquid coexistence region facilitates the choice
of the temperature and density for MD simulations in the liquid phase.
The aim is to reach low temperatures (and hence low vapour pressures)
while at the same time staying within the liquid region of the phase
diagram. The choice $\rho^{*}=0.27$ gives access to the liquid region of
the phase diagram. Moreover, for a typical molecular ionic liquid with
$\sigma = 5~\mbox{\AA}$ and ion mass $2m \simeq
150~\mbox{g}~\mbox{mol}^{-1}$, this density equates to a mass density of
about $1~\mbox{g}~\mbox{cm}^{-3}$, which  closely corresponds to
experimental values. The temperature range $1.5 \leqslant T^{*} \leqslant 3.0$
will be considered, as the upper limit is well within the liquid region
for all models, while the lower limit approaches the region of spinodal
decomposition.

\section{Liquid-phase structure and dynamics} \label{sec:strdyn}

\subsection{Molecular dynamics simulations} \label{sec:md}

MD simulations were performed using the LAMMPS package~\cite{Plimpton:1995/a,url_lammps}. Systems of $N=125$ ion pairs were
simulated in a cubic box with edge length $L=7.736\sigma$ giving a
reduced ion-pair density of $\rho^{*}=0.27$, as advertised in section~\ref{sec:vl}. The dynamical equations of motion were integrated using
the velocity-Verlet scheme with a timestep of $\delta t^{*}=0.0025$.
Simulations were started from a CsCl lattice structure and equilibrated
at the desired temperature using simple velocity rescaling over $10^{5}$
timesteps. Production runs were then performed in the canonical ($NVT$)
ensemble over $7.2 \times 10^{6}$ timesteps (equivalent to around
$35~\mbox{ns}$) with weak coupling to a Berendsen thermostat~\cite{Allen:1987/a}.
%
%
\begin{table}[htb]
\caption{\label{tab:pressure} Reduced pressure $P^{*}$ along the
isochore $\rho^{*}=0.27$ as a function of temperature, for each of the
charged models.}
\vspace{2ex}
\centering
\begin{tabular}{|l|r|r|r|r|r|} \hline
& \multicolumn{5}{c|}{Model} \\ \hline
$T^{*}$ &      SD &      AD &      SS &      AS &      CS \\ \hline\hline
$1.50$  & $-1.76$ & $-1.62$ & $-1.78$ & $-0.30$ & $-1.52$ \\
$1.75$  & $-1.13$ & $-1.36$ &  $0.16$ &  $0.41$ &  $0.08$ \\
$2.00$  & $-0.45$ & $-0.75$ &  $1.17$ &  $1.10$ &  $1.17$ \\
$2.25$  &  $0.22$ & $-0.11$ &  $1.85$ &  $1.76$ &  $1.85$ \\
$2.50$  &  $0.88$ &  $0.53$ &  $2.48$ &  $2.40$ &  $2.49$ \\
$2.75$  &  $1.53$ &  $1.16$ &  $3.12$ &  $3.02$ &  $3.13$ \\
$3.00$  &  $2.17$ &  $1.79$ &  $3.73$ &  $3.63$ &  $3.73$ \\ \hline
\end{tabular}
\end{table}
Ion positions and velocities, and all components of
the stress tensor, were saved at intervals of 4 timesteps for
post-processing. The production runs were divided into 12 blocks, and
statistical uncertainties were estimated by assuming block averages to
be statistically independent. Table~\ref{tab:pressure} shows the values
of the pressure along the isochore $\rho^{*}=0.27$ for each of the
charged models. At temperatures $T^{*} < 2.5$, the pressure in at least
one of the systems is negative, indicating that these state points may
well be mechanically unstable and/or within a binodal region.
Figure~\ref{fig:vl} confirms that these low temperatures are approaching the
vapour-liquid coexistence region.

\subsection{Structure} \label{sec:rdfs}

\begin{figure}[!h]
\centering
\includegraphics[width=10cm]{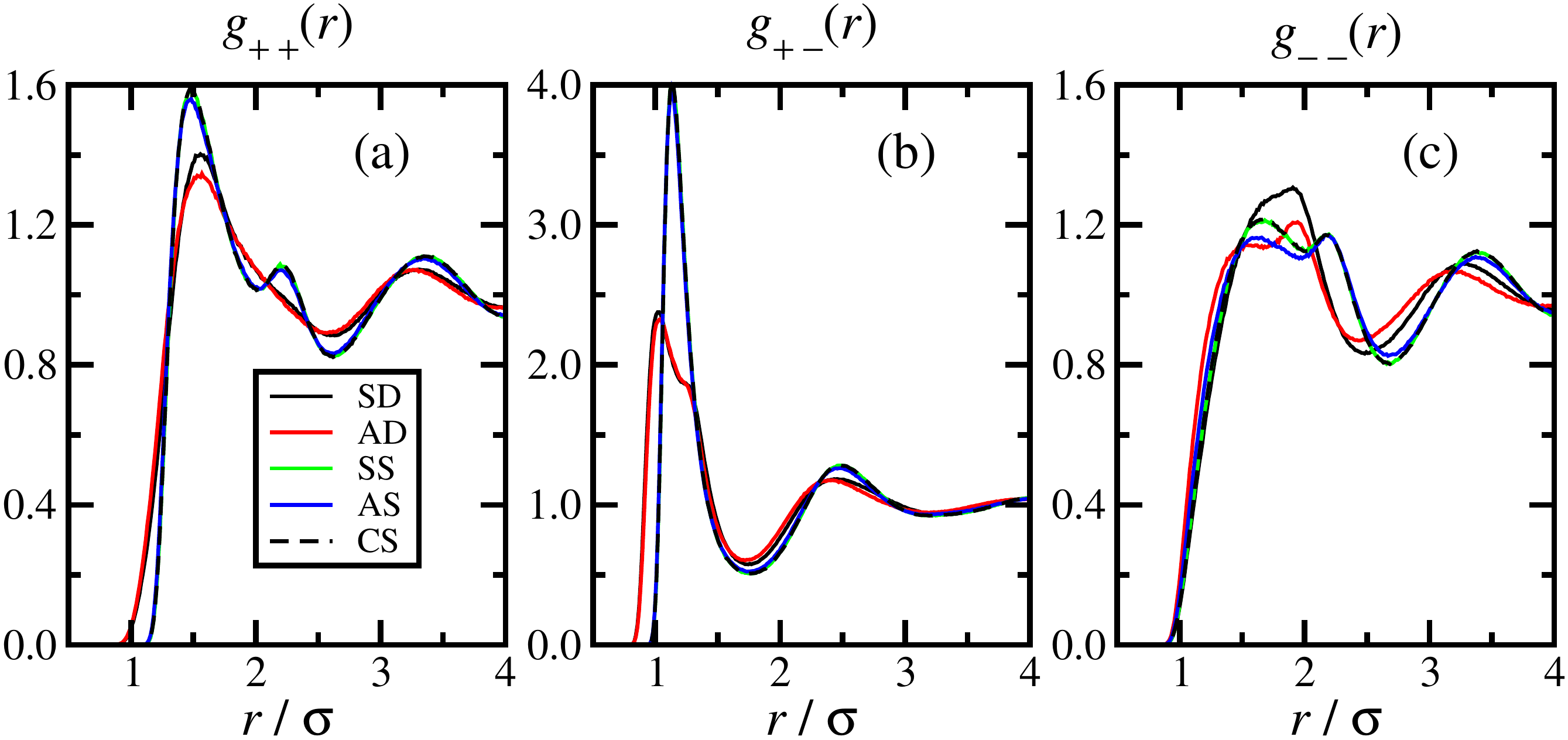}
\caption{\label{fig:correlations} (Color on-line) Pair-correlation functions in the
liquid phase at $\rho^{*}=0.27$ and $T^{*}=2.5$: (a) cation-cation RDF,
$g_{++}(r)$; (b) cation-anion RDF, $g_{+-}(r)$; (c) anion-anion RDF,
$g_{--}(r)$. In all cases, the separation $r$ is that between centres of
mass.}
\end{figure}

The microscopic structure in the liquids was examined using the radial
distribution functions (RDFs) $g_{\alpha\beta}(r)$ ($\alpha,\beta=+,-$)~\cite{Hansen:2006/a}. These functions are given by
\begin{equation}
g_{\alpha\beta}(r) = \frac{V}{4\pi N_{\alpha}N_{\beta} r^{2}}
                     \left\langle
                      \sum_{i}^{N_{\alpha}}
                      \sum_{j\ne i}^{N_{\beta}}
                      \delta{(r-r_{ij})}
                     \right\rangle ,
\end{equation}
where $N_{+}=N_{-}=N$, and $r_{ij}$ is the distance between the
centres of mass of particles $i$ and $j$. MD simulation results at
$T^{*}=2.5$ are shown in figure~\ref{fig:correlations}; this
temperature was selected in order to be sure of simulating a
single dense liquid phase.
%
%
Firstly, the cation-cation RDF
($g_{++}$) and anion-anion RDF ($g_{--}$) show primary peaks at
around $r \simeq 1.5\sigma$ for all models. The cation-anion RDF
($g_{+-}$) shows strong correlations at much shorter distances,
but there is a marked difference between the spherical models (SS,
AS, CS) and the dumbbell models (SD, AD), reflecting the shapes of
the cations. The spherical models show one primary peak at $r
\simeq 1.1\sigma$, roughly equal to the distance of closest
approach $(\sigma+\sigma_{+})/2 \simeq 1.15\sigma$. The dumbbells
models exhibit a short-range double-peaked feature: the peak near
$r \simeq 1\sigma$ is clearly due to the anion sitting in the
`valley' between the two spheres of the cation dumbbell; the peak
near $r \simeq 1.3\sigma$ roughly corresponds to the anion being
located near the dumbbell axis, at a distance $\sigma+d/2 \simeq
1.21\sigma$ from the cation centre of mass. This picture is in
good correspondence with work on similar models by Malvaldi and
Chiappe~\cite{Malvaldi:2008/a}.

\subsection{Dynamics} \label{sec:dynamics}

\begin{figure}[tb]
\centering
\includegraphics[width=12cm]{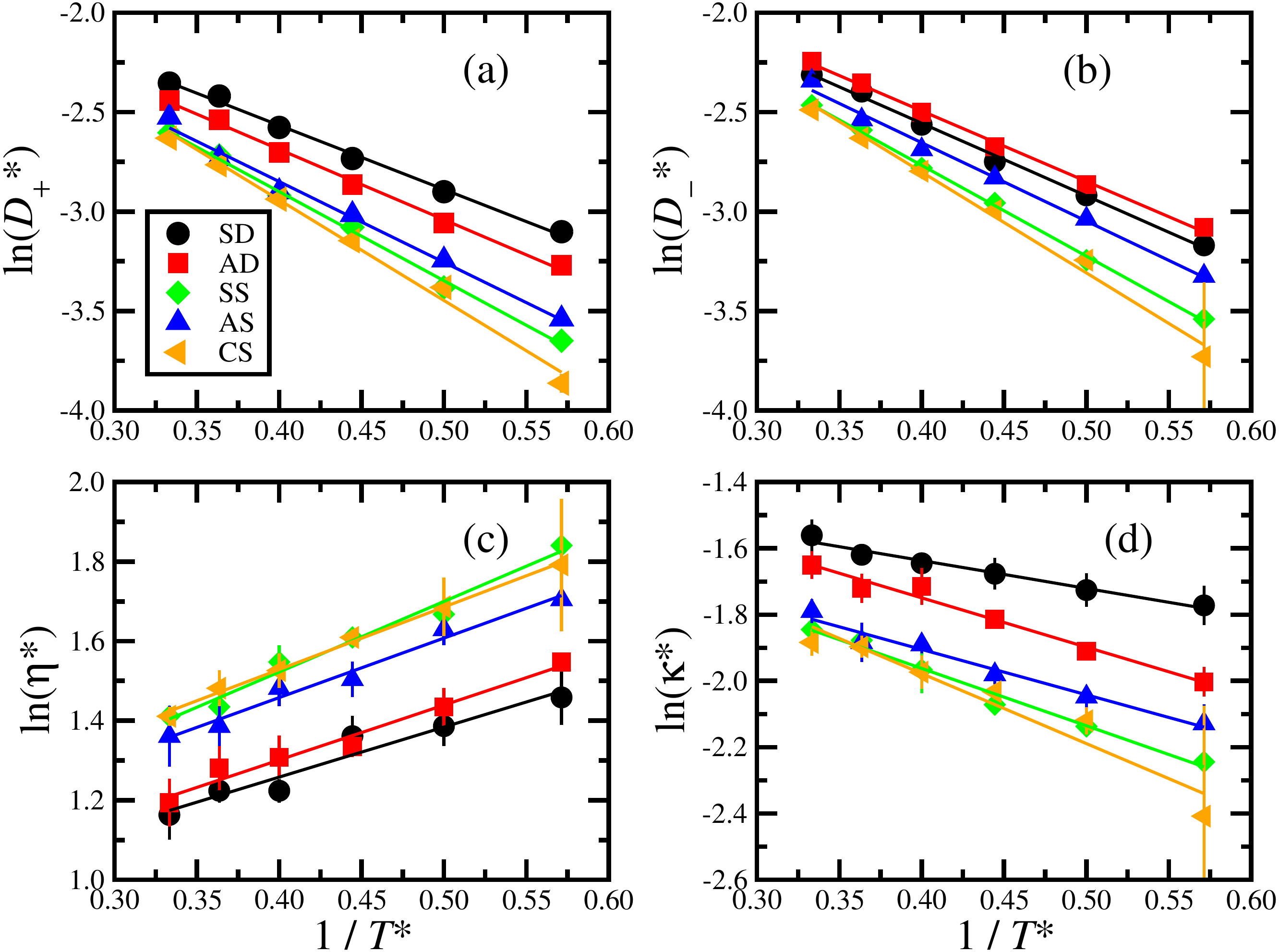}
\caption{\label{fig:arrhenius}(Color on-line) Arrhenius plots of transport properties
at $\rho^{*}=0.27$ and $1.75 \leqslant T^{*} \leqslant 3.00$: (a) cation diffusion
constant $D_{+}$; (b) anion diffusion constant $D_{-}$; (c) shear
viscosity $\eta$; (d) static conductivity $\kappa$. Results are plotted
in reduced units defined in the text.}
\end{figure}

A range of dynamical properties has been calculated. The
frequency-dependent conductivity, $\kappa(\omega)$, is related to the
charge current defined by
\begin{equation}
{\bf J}(t) = \sum_{i=1}^{N_{q}}
             q_{i}
             \left[{\bf v}_{i}^{\rm com}(t)+\Delta{\bf v}_{i}(t)\right],
\end{equation}
where the sum runs over all charge sites (of which there are $N_{q}$),
$\Delta{\bf v}_{i}(t)$ is the velocity of charge $i$ in its molecular
centre-of-mass frame, and ${\bf v}_{i}^{\rm com}(t)$ is the
corresponding molecular centre-of-mass velocity. (Obviously, $\Delta{\bf
v}=0$ for the anions.) The frequency-dependent conductivity is given by~\cite{Hansen:2006/a}
\begin{equation}
\kappa(\omega)
= \kappa'(\omega) + {\rm i}\kappa''(\omega)
= \frac{1}{3Vk_{\rm B}T}
  \int_{0}^{\infty}
  \langle{\bf J}(t)\cdot{\bf J}(0)\rangle
  \exp{(-{\rm i}\omega t)}
  {\rm d}t,
\end{equation}
where ${\rm i}=\sqrt{-1}$, and $\kappa'(\omega)$ and $\kappa''(\omega)$
are the real (in-phase) and imaginary (out-of-phase) parts,
respectively. The static conductivity is $\kappa=\kappa(0)$. The
contributions of different single-particle motions to the conductivity
spectrum can be determined by calculating ion translational velocity
autocorrelation functions $C_{v}(t)$~\cite{Hansen:2006/a}, and a cation
intramolecular velocity autocorrelation function $C_{\Delta}(t)$:
\begin{eqnarray}
C_{v}(t)
&=& \langle{\bf v}_{i}^{\rm com}(t)\cdot{\bf v}_{i}^{\rm com}(0)\rangle, \\
C_{\Delta}(t)
&=& \langle\Delta{\bf v}_{i}(t)\cdot\Delta{\bf v}_{i}(0)\rangle.
\end{eqnarray}
These two functions capture the dynamics of the molecular centre of mass
[$C_{v}(t)$] and the intramolecular rotations [$C_{\Delta}(t)$]~\cite{Schroder:2009/a}. The ion diffusion constants are obtained from
the standard Green-Kubo relationship
\begin{equation}
D = \frac{1}{3}\int_{0}^{\infty}C_{v}(t){\rm d}t.
\end{equation}
The reorientational dynamics of the cations are examined by calculating
the correlation function
\begin{equation}
C_{e}(t) = \langle{\bf e}_{i}(t)\cdot{\bf e}_{i}(0)\rangle,
\end{equation}
where ${\bf e}(t)$ is the orientation unit vector along the cylindrical
symmetry axis of the cation. An associated correlation time can be
defined by
\begin{equation}
\tau_{e} = \int_{0}^{\infty}C_{e}(t){\rm d}t.
\end{equation}
Finally, the viscosity of the fluid is calculated using the Green-Kubo
formula involving the autocorrelation function of the off-diagonal
elements of the stress tensor $\Pi_{xy}(t)$:
\begin{equation}
\eta = \frac{1}{Vk_{\rm B}T}\int_{0}^{\infty}
       \langle \Pi_{xy}(t)\Pi_{xy}(0) \rangle {\rm d}t.
\end{equation}
The static transport properties are considered first, namely the
diffusion constants ($D$), shear viscosity ($\eta$), and static
conductivity ($\kappa$). In reduced units, these properties are defined
by $D^{*}=D\tau/\sigma^{2}$, $\eta^{*}=\eta\sigma^{3}/\epsilon\tau$, and
$\kappa^{*}=\kappa\tau/{\cal E}$. These properties were measured for
each charged model as functions of temperature. In accord with Eyring's
activated-dynamics picture~\cite{Eyring:1936/a}, it turns out that the
data are fitted adequately with Arrhenius relations of the form
\begin{equation}
D      \propto \exp{(-E_{\rm a}/k_{\rm B}T)}, \qquad \eta \propto
\exp{( E_{\rm a}/k_{\rm B}T)}, \qquad \kappa \propto \exp{(-E_{\rm
a}/k_{\rm B}T)},
\end{equation}
where $E_{\rm a}$ is an activation energy associated with that
particular transport property. The logarithms of the transport
properties are plotted as functions of $1/T^{*}$ in
figure~\ref{fig:arrhenius}, along with Arrhenius-law fits.
Simulations at $T^{*}=1.5$ gave erratic and anomalous results,
probably due to the system being within either the liquid-solid or
vapour-liquid coexistence region; these data points are omitted
from the plots. Arrhenius-like behaviour is observed over the
temperature range considered. In general, for a given temperature,
the cation and anion diffusion constants obey $D_{+} < D_{-}$ due
to the larger steric bulk of the cations. The dumbbells models (SD
and AD) exhibit higher values of $D_{+}$ than the sphere models
(SS, AS, CS). This may be attributed to the elongated shape of the
dumbbell cations, facilitating the motion along their long axes.
For the dumbbell cations, a symmetric charge distribution gives a higher value of
$D_{+}$ than an asymmetric charge distribution, which
probably arises from there being less strong interactions (on
average) between an SD cation and its nearest-neighbour anions.
For the spherical cations, the reverse is true, although the
magnitude of the effect is smaller. In general, $D_{-}$ mirrors
the trends seen in $D_{+}$\,, indicating strong association
between oppositely charged ions. Using the molecular parameters
given in section~\ref{sec:models}, the reduced values $D^{*}
\simeq 0.03$--$0.10$ correspond to real values $D \simeq
0.3$--$1.0 \times 10^{-7}~\mbox{m}^{2}~\mbox{s}^{-1}$, which are a
little higher than typical results for RTILs. The shear viscosity
$\eta$ shows the reverse trend to $D_{\pm}$\,, with the dumbbells
models being least viscous; this can be rationalised qualitatively
on the basis of a Stokes-Einstein relationship of the form $D
\propto 1 / \eta$~\cite{Hansen:2006/a}. The reduced values
$\eta^{*} \simeq 3$--$5$ equate to real values of $\eta \simeq
3$--$5 \times 10^{-4}~\mbox{Pa}~\mbox{s}$. These are much smaller
than the viscosities of real RTILs, which are normally on the
scale of tens of $\mbox{mPa}~\mbox{s}$. The trends in the static
conductivity mirror those in $D_{\pm}$\,, which can be
rationalised with a generalised Nernst-Einstein relationship of
the form~\cite{Hansen:2006/a}
\begin{equation}
\kappa = \frac{\rho q^{2}}{k_{\rm B}T}\left( D_{+} + D_{-} \right)
         \left( 1 - \Delta \right),
\label{eqn:ne}
\end{equation}
where $\Delta$ characterises the deviation from pure
Nernst-Einstein behaviour, which is observed when there are no
cross correlations between cation and anion motions. Such
correlations lead to a reduction in $\kappa$ and hence $\Delta >
0$. For the data shown in figure~\ref{fig:arrhenius}, $\Delta
\simeq 0.75$, a large value that arises from very strong
cation-anion correlations. The reduced values $\kappa^{*} \simeq
0.1$--$0.2$ correspond to real values of $\kappa \simeq
6$--$12$~S~m$^{-1}$, which are somewhat higher than
those seen in experiments on RTILs. In summary, then, for dumbbell
cations, a symmetric charge distribution gives higher diffusion
constants, lower shear viscosity, and higher static conductivity.
The dependences of transport properties with spherical cations are
less sensitive to the charge distribution.


The activation energies associated with the static transport
properties are reported in table~\ref{tab:activation}. In general,
the dumbbell-cation models exhibit lower activation energies. This
correlates with the picture presented above, with the elongated
models being more capable of breaking out of transient molecular
cages by moving along their long axes. When converted to real
energy units~-- using the molecular parameters given in
section~\ref{sec:models}~-- the activation energies for diffusion
are in the region of $20~\mbox{kJ}~\mbox{mol}^{-1}$, which is a
typical value for some RTILs~\cite{Hayamizu:2010/a}. The
activation energies for viscosity and conductivity are in the
region of $5$--$10~\mbox{kJ}~\mbox{mol}^{-1}$, which are the right
order of magnitude but a little too low. Experimental results for
RTILs show significant deviations from Arrhenius behaviour and are
usually fitted with a Vogel-Fulcher-Tammann equation of the form
$A\exp{[B/(T-T_{0})]}$, so these are not directly comparable to
the Arrhenius-law fits presented here. Nonetheless, $B/k_{\rm B}$
is typically of order $500$--$1000~\mbox{K}$, which in the current
reduced units equates to around $B/\epsilon \simeq 1$--$2$; this
is at least the same order of magnitude as the activation energies
presented in table~\ref{tab:activation}.


\begin{table}[!t]
\caption{\label{tab:activation} Activation energies obtained from
Arrhenius-law fits to the temperature dependence of the cation diffusion
constant ($D_{+}$), the anion diffusion constant ($D_{-}$), the shear
viscosity ($\eta$), the static conductivity ($\kappa$), and an
orientational correlation time ($\tau_{e}$). Figures in brackets are
estimated uncertainties in the final digits.}
\vspace{2ex}
\centering
\begin{tabular}{|l|l|l|l|l|l|}\hline
& \multicolumn{5}{c|}{Activation energies $E_{\rm a}/\epsilon$} \\ \hline
Model & $D_{+}$ & $D_{-}$ & $\eta$ & $\kappa$ & $\tau_{e}$ \\ \hline\hline
SD & $3.28(13)$ & $3.71(13)$  & $1.27(21)$  & $0.823(80)$ & $2.452(43)$ \\
AD & $3.63(12)$ & $3.636(96)$ & $1.47(10)$  & $1.532(93)$ & $2.938(51)$ \\
SS & $4.51(14)$ & $4.580(89)$ & $1.806(73)$ & $1.70(14)$  & $1.301(39)$ \\
AS & $3.98(26)$ & $4.10(25)$  & $1.55(14)$  & $1.47(11)$  & $1.748(37)$ \\
CS & $4.79(18)$ & $4.546(38)$ & $1.725(62)$ & $1.51(12)$  & \\ \hline
\end{tabular}
\end{table}


\begin{figure}[!b]
\centerline{
\includegraphics[width=8cm]{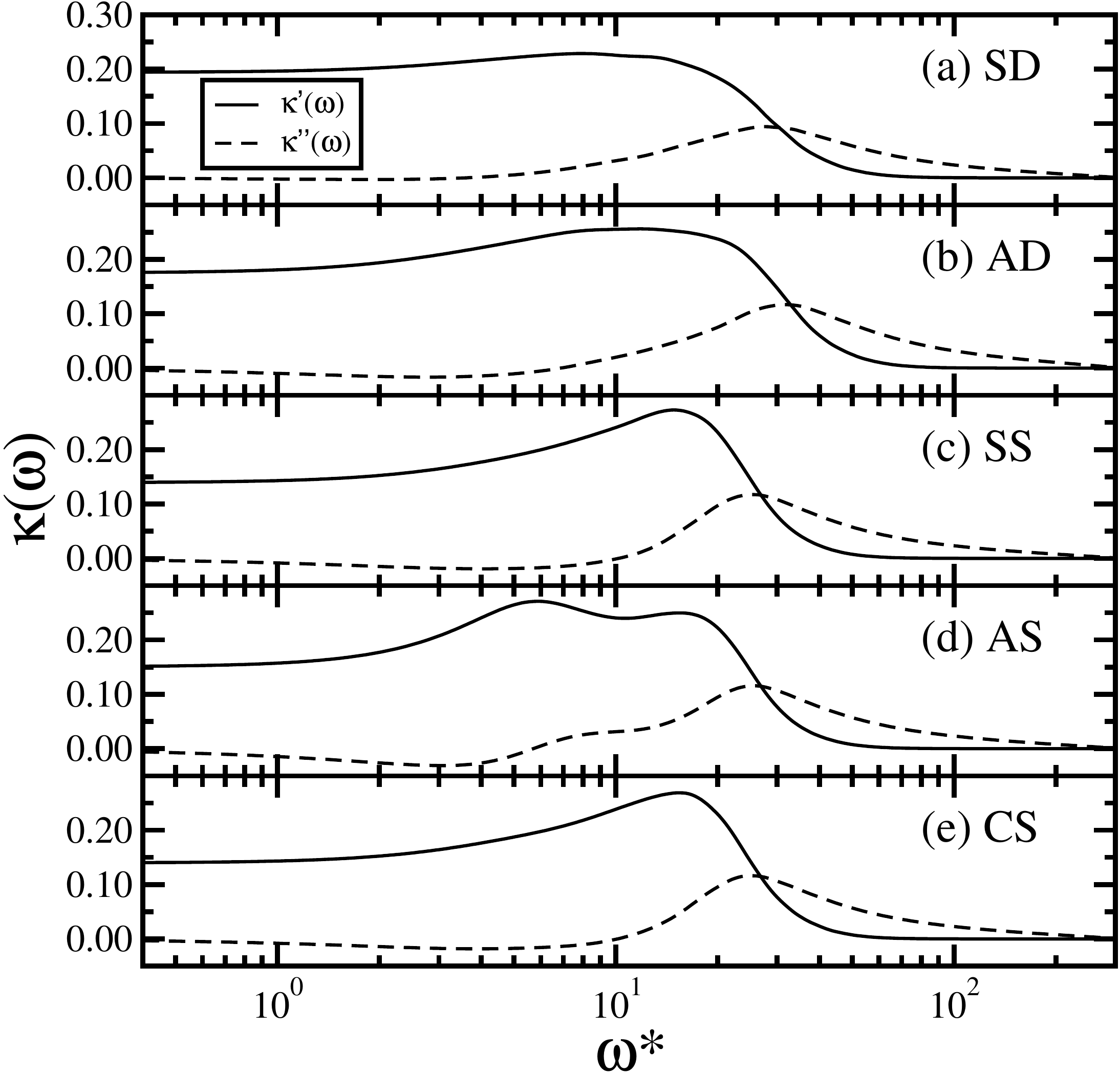}}
\caption{\label{fig:conductivity} Real and imaginary parts of the
conductivity spectrum, $\kappa'(\omega)$ (solid lines) and
$\kappa''(\omega)$ (dashed lines), respectively, for primitive models at
$\rho^{*}=0.27$ and $T^{*}=2.5$: (a) SD model; (b) AD model; (c) SS
model; (d) AS model; (e) CS model.}
\end{figure}
The dynamics are now considered in more detail for a single state
point, $\rho^{*}=0.27$ and $T^{*}=2.5$. This temperature is chosen
such that each of the models is in its liquid phase, away from the
coexistence region. Figure~\ref{fig:conductivity} shows the
conductivity spectra $\kappa(\omega)$ for all five models. The
spectra of the SD and AD models are characterised by broad
shoulders in the real parts, $\kappa'(\omega)$, at frequencies
$\omega^{*} \simeq 10$--$20$, and single peaks in the imaginary
parts, $\kappa''(\omega)$, at a frequency $\omega^{*} \simeq 30$.
The spectrum for the CS model is particularly simple, with the
real part exhibiting a peak that is characteristic of ionic
liquids~\cite{Hansen:1975/a,Svishchev:1994/a,Petravic:2003/a}. The
low-frequency, negative portion of the imaginary part signals that
ion translations may contribute to the low-frequency complex
dielectric spectrum $\varepsilon(\omega)=1+4\pi{\rm
i}\kappa(\omega)/\omega$~\cite{Shim:2008/a}. The spectrum of the
SS model looks very similar to that of the CS model, showing that
a symmetrical displacement of charge on the cation does not
strongly affect the conductivity. This is natural, because these
two models share the same centre of mass and the same centre of
charge; the displacement of charge should only affect the degree
of interactions between cations and anions, and is not expected to
lead to new qualitative features in $\kappa(\omega)$. Very
interesting features arise in the spectrum of the AS model. Here,
the real part contains two, very-well resolved peaks at reduced
frequencies of around $\omega^{*}=5$ and $\omega^{*}=20$.


\begin{figure}[!b]
\centering
\includegraphics[width=10cm]{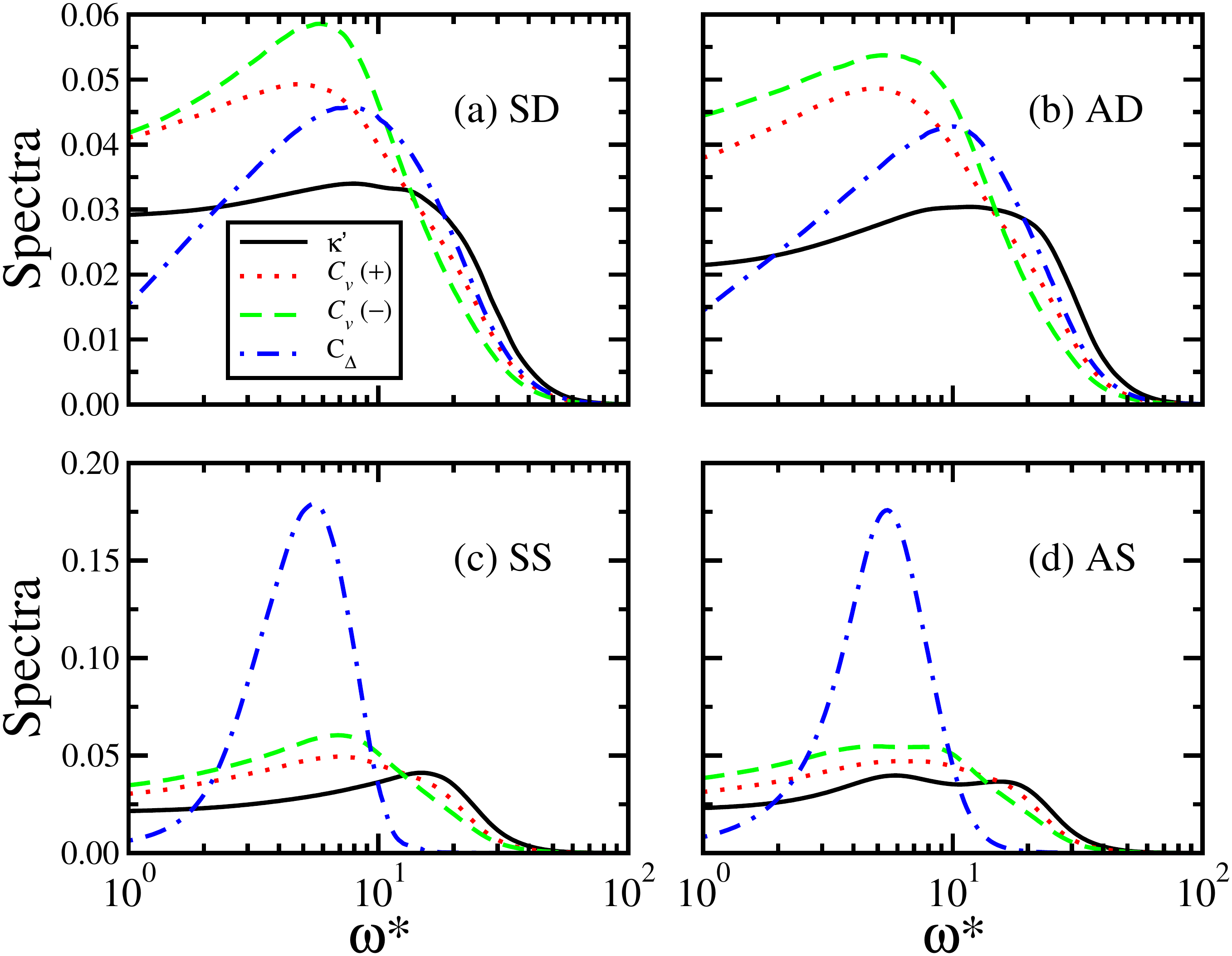}
\caption{\label{fig:spectra} (Color on-line) Real part of the conductivity
spectrum [$\kappa'(\omega)$] compared with the spectra of the
centre-of-mass velocity autocorrelation functions for the cations
and anions [$C_{v}(\omega)$], and the rotational velocity
autocorrelation function for the cations [$C_{\Delta}(\omega)$],
for primitive models at $\rho^{*}=0.27$ and $T^{*}=2.5$: (a) SD
model; (b) AD model; (c) SS model; (d) AS model. In each case, the
plots are normalised so that the frequency integral for each
spectrum is equal to unity.}
\end{figure}

To help identify the molecular motions responsible for the
features in $\kappa(\omega)$, figure~\ref{fig:spectra} shows the
real part along with the spectra of $C_{v}(t)$ (for each of the
ions) and that of $C_{\Delta}(t)$:
\begin{equation}
C(\omega) = 2\int_{0}^{\infty}C(t)\cos{(\omega t)}{\rm d}t.
\end{equation}
Recall that $C_{v}(t)$ describes the centre-of-mass motions of the ions,
while intramolecular rotations of the cations are described by
$C_{\Delta}(t)$. The spectra are basically densities of states for
vibrations and librations within the solvation shell. Of course, these
are single-particle properties, and in the following discussion, the
cross-correlations between cations and anions (that lead to deviations
from the Nernst-Einstein law) are ignored. Nonetheless, the
single-particle properties will yield some insight on the collective
dynamics that lead to the conductivity spectrum. For the purposes of
comparison, each spectrum is normalised to unit area. For the SD and AD
models, the cation translations, anion translations, and cation
rotations occur on similar timescales. As a result, $\kappa'(\omega)$
exhibits a broad peak over the relevant range of frequencies. For the SS
and AS models, the rotational spectra are much less broad than those for
the SD and AD models. For the AS model, the peak in the rotational
spectrum coincides with the low-frequency peak in $\kappa'(\omega)$,
indicating that this feature of the conductivity arises from cation
rotations. Of course, this feature is absent from $\kappa'(\omega)$ for
the SS model, because the rotation of a symmetrical distribution of
positive ions leads to no net transport of charge.

The peaks in the rotational spectra in figure~\ref{fig:spectra}
show that the spherical cations (SS and AS) librate slower than
the dumbbell cations, despite their having the same excluded
volumes and equal moments of inertia. This difference is
attributed to the so-called charge arm, described by Kobrak and
Sandalow~\cite{Kobrak:2006/a}. If a single charge $q$ is displaced
by a distance $l_{q}$ from the cation centre of mass, then the
librational frequency of rotation is proportional to
$l_{q}/I$~\cite{Kobrak:2006/a}; $l_{q}$ is called the charge arm.
Recall that in the present models, the charges are set a certain
distance from the edge of the cations, and so they may be at
different distances from the centres of mass. For the dumbbell
models $l_{q}=\sigma/2\sqrt{2} \simeq 0.3536\sigma$, while for the
spherical models $l_{q}=0.1507\sigma$. Keeping the moment inertia
$I^{*}=0.25$ for the AD model and equating the ratio $l_{q}/I$ for
the AD and AS models gives $I^{*}=0.1066$ for the AS model.
Conductivity spectra and the spectra of the velocity
autocorrelation functions have been calculated for the SS and AS
models with $I^{*}=0.1066$ (data not shown) and the rotational
peak does indeed shift to higher frequency, coinciding with the
peak frequencies for the SD and AD models. Correspondingly, the
low-frequency peak in $\kappa'(\omega)$ shifts up in frequency and
merges with the centre-of-mass peak.

The single-particle motions detailed here occur in the region of
$\omega^{*} \simeq 5$--$20$, which equate to real frequencies $\nu
\simeq 0.4$--$1.6~\mbox{THz}$. These are quite close to the
natural timescales in RTILs: cation rotations in
1-alkyl-3-methylimidazolium salts occur on the timescale of tens
of picoseconds~\cite{Schroder:2007/a,Wulf:2007/a,Huang:2010/a}.
Shim and Kim present $\kappa(\omega)$ from computer simulations of
1-ethyl-3-methylimidazolium hexafluorophosphate that exhibits a
main peak at around $15~\mbox{THz}$ and a low-frequency shoulder
below about $10~\mbox{THz}$~\cite{Shim:2008/a}; it is tempting to
speculate that the low-frequency feature arises from cation
rotations.


\begin{figure}[!h]
\centering
\includegraphics[width=10cm]{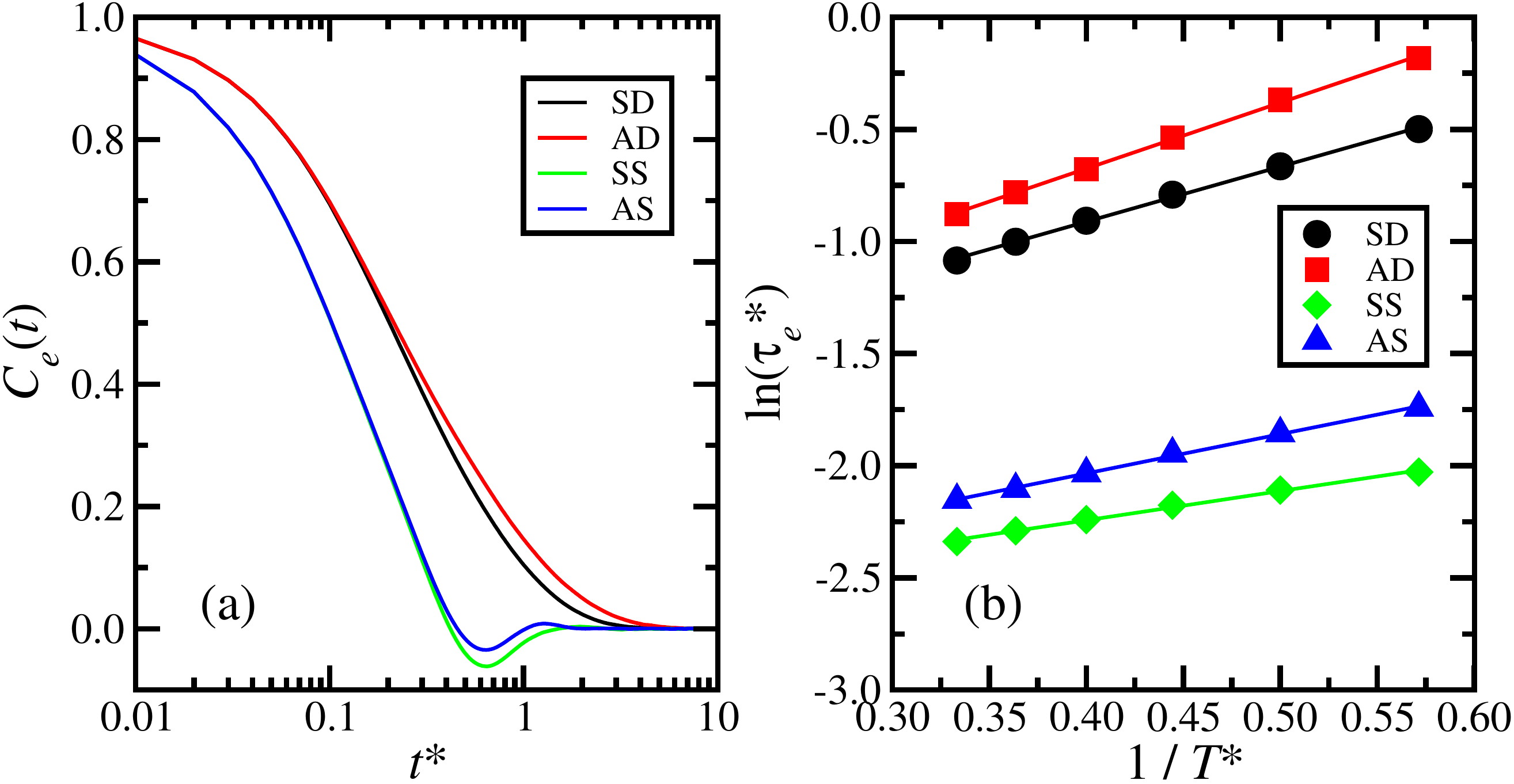}
\caption{\label{fig:oacf}(Color on-line) (a) Orientational autocorrelation function
$C_{e}(t)$ for primitive models at $\rho^{*}=0.27$ and $T^{*}=2.5$. (b)
Arrhenius plot of the corresponding decay times $\tau_{e}$ for $1.75
\leqslant T^{*} \leqslant 3.00$. Results are plotted in reduced units defined in
the text.}
\end{figure}

To explore cation reorientations further, figure
\ref{fig:oacf}~(a) shows $C_{e}(t)$ for the dumbbell-cation and
spherical-cation models, at $\rho^{*}=0.27$ and $T^{*}=2.5$.
Firstly, the dumbbell models exhibit a monotonous decay in
$C_{e}(t)$, while the spherical models show a negative portion.
The spherical models may undergo a rapid reversal in orientation
on the timescale of $t \simeq 0.6\tau$ due to the absence of steric
hindrance. Secondly, the correlations die away more slowly with
dumbbells than with spheres, due to the packing effects that
restrict the reorientation of the anisotropic dumbbells. Finally,
an asymmetric charge distribution, as in the AD and AS models,
leads to a slower decay of $C_{e}(t)$ as compared to the SD and SS
models, respectively. This may be due to the anisotropy of the
cation-anion interactions when a single, whole charge is localised
near the edge of the cation, rather than being spread out over the
molecule. To summarise, anisotropies in the short-range repulsions
and electrostatic interactions lead to longer orientational
correlation times. The decay time $\tau_{e}^{*} = \tau_{e}/\tau$
has been calculated as a function of temperature for each of the
models; the results are presented on Arrhenius plots in
figure~\ref{fig:oacf}~(b). The associated activation energies from
the Arrhenius law $\tau_{e} \propto \exp{(E_{\rm a}/k_{\rm B}T)}$
are reported in table~\ref{tab:activation}. The activation
energies for the dumbbell models are greater than those for the
spherical models, once again reflecting the steric barriers to
rotation of dumbbells arising from their anisotropic shapes. An
asymmetric charge distribution gives rise to marginally higher
barriers than does a symmetric one, reflecting the effect of
anisotropy of cation-anion interactions. The decay times span the
range $\tau_{e}^{*} \simeq 0.1$--$0.8$, which in real units
correspond to the picosecond timescale; these are realistic
values, at least for simple molecular liquids. When converted to
real units, the activation energies are of order
$10~\mbox{kJ}~\mbox{mol}^{-1}$, which as mentioned above, are at
least of the right order of magnitude for molecular ionic liquids,
but a bit too low.

\section{Conclusions} \label{sec:conclusions}

Simple models of molecular ionic liquids were constructed in order to
study the effects of ion-size disparity, cation-charge distribution, and
cation-shape anisotropy. The cations were made up of charged soft
spheres, with attractive interactions included: one type of cation
consisted of a single sphere carrying either one charge at the centre or
off-centre, or two charges displaced symmetrically from the centre; the
other type of cation was a dumbbell, with either one or two of the
constituent spheres carrying a charge. The anions were charged soft
spheres. Such simple models admit in-depth studies of phase behaviour
and dynamics in simulations of reasonable length, using normal
computational facilities. The vapour-liquid coexistence curves were
determined using Wang-Landau simulations and the resulting critical
parameters were compared with those of the restricted primitive model,
the simplest representation of an ionic fluid. The presence of
attractive interactions increases the critical temperature above that
expected from electrostatic interactions alone, while the critical
density remains roughly constant and characteristically low.

Dynamics simulations were then conducted along a liquid-state isochore
appropriate to typical ambient conditions. The microscopic structure has
been studied using pair correlation functions, and rationalised in terms
of the probable packing of the ions, and the resulting electrostatic
interactions. The temperature dependences of the ion diffusion
constants, shear viscosity, conductivity, and orientational correlation
time are Arrhenius-like, at least over the temperature range considered
here. The magnitudes of these properties, and the associated activation
energies, when converted to real units, are not too dissimilar from
those measured experimentally, although in general the diffusion
constants and conductivity are too high, and the viscosity is too low.
This is to be expected of such simple models with comparatively little
`molecular roughness'. Coarse graining usually results in an effective
decrease of molecular-scale friction, implying enhanced diffusion and
conductivity, and reduced viscosity. This is a general drawback of
coarse-grained models~\cite{Peter:2010/a} which cannot always be
remedied just by tuning the effective interactions; for instance,
molecular-scale noise and friction can be restored with integration
schemes similar to those used in dissipative particle dynamics
simulations~\cite{Hijon:2010/a}. The variations in the transport
coefficients between different models were rationalised in terms of
microscopic correlations. The dynamics at a particular liquid-phase
state point (high density, low temperature) were investigated in more
detail by looking at the frequency-dependent conductivity, and the
spectra of centre-of-mass and rotational velocity autocorrelation
functions. For the models with dumbbell cations, the centre-of-mass and
rotational motions occur on similar timescales, although only the
centre-of-mass motions contribute to the conductivity spectrum if the
positive charge is distributed symmetrically over the molecule. For the
models with spherical cations and off-centre charges, the centre-of-mass
and rotational motions occur on different timescales which, with an
asymmetrical distribution of charge on the cation, lead to well-resolved
features in the conductivity spectrum. The decay of orientational
correlations was rationalised with reference to anisotropies of short-range repulsions and electrostatic interactions. Despite the
simplicity of the models studied here, it is possible that some of the
features observed in the detailed dynamics may also be observable in
experimental studies of molecular ionic liquids.

\section*{Acknowledgements}

The authors are privileged to contribute this work to the special issue
of Condensed Matter Physics in honour of Professor Yura Kalyuzhnyi's
$60^{\rm th}$ birthday.



\ukrainianpart

\title{Фазова поведінка та динаміка в примітивних моделях молекулярно-іонних рідин}

\author{Г.С.~Ганзенмюллер\refaddr{label1}, Ф.Дж.~Кемп\refaddr{label2}}

\addresses{\addr{label1} Інститут динаміки швидкотривалих процесів
товариства Фраунгофера (Інститут Ернста Маха), Фрайбург, Німеччина
 \addr{label2} Школа хімії, Університет Едінбурга, Едінбург, Великобританія}

\makeukrtitle

\begin{abstract}
\tolerance=3000%
Досліджено фазову поведінку і динаміку молекулярно-іонних рідин з допомогою
примітивних моделей та масштабних комп'ютерних моделювань. Використані моделі
враховують різницю в розмірах катіона та аніона, положення заряду на катіонах і
анізотропію форми катіона, що є визначальними властивостями іонних рідин
при кімнатних температурах. Високоточне моделювання методом Монте-Карло використано
для побудови фазових діаграм рідина-газ, які в подальшому використовуються для
вивчення динаміки іонних рідин. За допомогою моделювання молекулярною
динамікою досліджено структуру, трансляційні та обертальні автокореляційні
функції, катіонні орієнтаційні автокореляції, самодифузію, в'язкість та частотну
залежність провідності. Отримані результати виявляють низку молекулярних механізмів
переносу зарядів, включаючи молекулярну трансляцію, обертання та асоціацію.

\keywords іонні рідини, перехід рідина-газ, динаміка, комп'ютерне моделювання
\end{abstract}

\end{document}